\documentclass[12pt]{article}

    \usepackage[T1]{fontenc}
    \usepackage{graphicx,epsfig,amsmath,amsfonts,cite}
    \renewcommand{\abstract}{}
\begin{document}

\title{Adiabatic hydrodynamic modes in dielectric environment in random electric field}

\author{A. A. Stupka}
\maketitle
\begin{center} {\small\it Oles Honchar Dnipropetrovs'k National University, Gagarin
ave., 72, 49010 Dnipropetrovs'k, Ukraine\\antonstupka@mail.ru }
\end{center}


\begin{abstract}
Dielectric is considered in the electric field that has equal to
zero the first moment and different from zero the second moment of
strength in an equilibrium. Equations of ideal hydrodynamics are
obtained in such field for the case of the neglect of dissipative
effects. A new variable - second moment of electric field strength
 is included in Euler
equation.Temporal equation  for this variable is obtained on the
basis of Maxwell equations in the hydrodynamic approximation.
Adiabatic one-dimensional waves of small amplitude are studied in
this system. Proceed from the theoretical estimation of the
intracrystalline field in an ionic crystal the good consent of the
obtained numeral values of transversal velocity of this wave  with
transversal velocity of sound for izotropic crystals of alkaline
haloids is found.


\end{abstract}


\section*{Introduction}

We will produce description of dielectric, as a continuous
environment by the local quantities of density $\rho $, pressure
$P$
  and mass velocity  $\vec v$. Except these cleanly hydrodynamic quantities, usually in the
electrohydrodynamics a state of environment is characterized by
strength of electric field  $\vec E$
  (see \cite{[1]} and the literature over brought there).
We will consider  a not piezoelectric izotropic crystal. In a
crystal there is the microscopic electric field that is a random
quantity with equal to zero the first moment of strength  $\vec
E$. We will suppose that it is present the different from zero
second moment of electric field strength. As shown in
\cite{[3],[4]}, the external magnetic field with the indicated
statistical properties provides the shear resiliency of conducting
liquid at small oscillations. That is why interesting to study the
small long-wave oscillations of liquid dielectric in the random
electric field with the purpose of establishing a possible
connection with the shear resiliency of the random
intracrystalline field. Quadratic on an electric field the Maxwell
stress tensor is included in standard Euler equation \cite{[1]},
for that we will build temporal equation in the
electrohydrodynamic approximation taking into account statistical
properties of the field.


\section{Equations of ideal electrohydrodynamics in the random field}

We will write standard equations of ideal electrohydrodynamics,
ignoring all dissipative effects \cite{[1]}. Continuity equation
\begin{equation}\label{0} \partial _t \rho  + Div\rho \vec {v} = 0
\end{equation}
and Euler equation
\begin{equation}\label{2}
\partial _t \left( {\rho v_i } \right) +
\partial _k \pi _{ik}  = 0 ,
\end{equation}
denotations are here entered for a spatial derivative  $\partial
/\partial x_k =
\partial _k $
 and for the momentum stream density tensor
\begin{equation}\label{6} \pi _{ik}  = \rho v_i v_k  + P\delta
_{ik}  - \left( {\left\langle {E_i E_k } \right\rangle  -
\left\langle {E^2 } \right\rangle \delta _{ik} /2}
\right)\varepsilon /4\pi  .
\end{equation}
 Here  $\varepsilon $
  is relative permittivity that we will consider as a constant value, that allows to ignore striction pressure.

But if we consider fully ionized plasma or ionic liquid, electric
polarization corresponds to a rearrangement of the bound electrons
in the material, which creates an additional charge density, known
as the bound charge density. Then, all charge: ionic $\sigma$ and
bound $\sigma^b$, must be included to Euler equation, and instead
using dielectric induction $Div \vec D= \varepsilon Div \vec E
=4\pi \sigma $, we must use $Div \vec E=4\pi (\sigma+ \sigma^b)$.
That gives instead (\ref{6})
\begin{equation}\label{6} \pi _{ik}  = \rho v_i v_k  + P\delta
_{ik}  - \left( {\left\langle {E_i E_k } \right\rangle  -
\left\langle {E^2 } \right\rangle \delta _{ik} /2} \right)/4\pi  .
\end{equation}

  As we see, the Maxwell stress tensor that is fully determined by the second moment of electric field
  $\left\langle {E_i E_k } \right\rangle $
 is included in Euler equation. Brackets mean averaging on the random phases of electric field (see \cite{[5]} p. 439).
 We will study an izotropic dielectric, then the equilibrium value of the field correlation is such:

\begin{equation}\label{6b}
\left\langle {E_l E_m } \right\rangle _0  = \left\langle {E_0^2 }
\right\rangle \delta _{lm} /3 = const .
\end{equation}
 For closing of the system of equations
it is needed to build equation for this new variable. It is done
in analogy to the case of magnetic hydrodynamics in the random
external magnetic field \cite{[3],[4]}. We will write down Maxwell
equation for electric field as random quantity

\begin{equation}\label{4}
\partial \varepsilon \vec E/\partial t = c Rot {\vec B} - 4\pi q\vec v
.
\end{equation}
 Here $ q
$
  is a local density of charge. Because of electroneutrality it has an equilibrium value
  $
q_0  = 0 $. We will be interested in one-dimensional linear
adiabatic oscillations in this system. Temporal equation for a
central moment  $\left\langle {E_i E_k } \right\rangle $
 we will obtain from (\ref{4}), multiplying on $E_k $
  in the same spatio-temporal point and making symmetrization \cite{[3],[4]}. We will take into account
  that the equilibrium field $\vec E$
  exists in the immobile system, that is why, according to Galilean transformations \cite{[11]},
  ${\vec B'} = {\vec B} + [{\vec v},{\vec E}]\varepsilon /c$. Then, we can write down
\[\partial \varepsilon E_i E_k /\partial t = c Rot _i \left( {{ \vec{B}} + [{\vec v},{\vec E}]\varepsilon /c}
 \right)E_k  + c Rot _k \left( {{\vec B} + [{\vec v},{\vec E}]\varepsilon /c} \right)E_i-\]
\begin{equation}\label{4a}
  - 4\pi
  q\left( {v_i E_k  + v_k E_i } \right)
.
\end{equation}
 We consider the fields of velocity and magnetic induction
uncorrelated with electric one. After averaging on random phases
(see, for example, \cite{[5]} p. 439) and using (\ref{6b})   we
have the linearized equation

\begin{equation}\label{7}
\partial \left\langle {E_i E_k } \right\rangle /\partial t = \left( {\partial _k \vec v_i^{}  + \partial _i \vec v_k^{}  - 2\partial _l \vec v_l^{} \delta _{ik} } \right)\left\langle {E_0^2 } \right\rangle /3
,
\end{equation}
 Thermal fluctuations we ignored. The entropy
deviation is included in the momentum stream density tensor
(\ref{6}). We will consider an adiabatic process, then after
linearization we have for pressure $P = \left( {\partial
P/\partial \rho } \right)_s \rho $
  and (\ref{2}) looks as
\begin{equation}\label{8}
\partial _t v_i  + \partial _k \left\{ {\delta _{ik} v_s^2 \rho -
\left( {\left\langle {E_i E_k } \right\rangle  - \left\langle {E_l
E_l } \right\rangle \delta _{ik} /2} \right) /4\pi \rho _0 }
\right\} = 0 .
\end{equation}
 Here $\rho _0 $
  is an equilibrium value of mass density and
  $v_s^2  = \left( {\partial P/\partial \rho } \right)_s /\rho _0 $. From (\ref{0}) after linearization we have
\begin{equation}\label{9}
\partial _t \rho  + \rho
_0 Div\vec v = 0.
\end{equation}

\section{Adiabatic one-dimension waves of small amplitude}

We will consider one-dimensional waves, along direction of
distribution we will direct a coordinate axis  $z$. Let all
electrohydrodynamic quantities depend only on $z$
  and t. Equation (\ref{7}) is symmetric on tensor indexes  $i$
 and $k$
 that is why it contains 6 equations for the components of symmetric tensor
 $\left\langle {E_i E_k } \right\rangle $. Equation  $\partial _t \left\langle {E_x E_y } \right\rangle  =
0$
 has a trivial solution  $\left\langle {E_x E_y } \right\rangle  = const$
  and moves away from other system. Other 9 equations (\ref{7})-(\ref{9}) it comfortably to present in a matrix form
\begin{equation}\label{10} \partial _t \Psi _\alpha   + {\rm
Z}_{\alpha \beta } \partial _z \Psi _\beta   = 0.
\end{equation}
 A vector of state is here entered
 \begin{equation}\label{11}
\Psi  = \left( {\rho ,v_x ,v_y ,v_z ,\left\langle {E_x E_x }
\right\rangle ,\left\langle {E_x E_z } \right\rangle ,\left\langle
{E_y E_y } \right\rangle ,\left\langle {E_y E_z } \right\rangle
,\left\langle {E_z E_z } \right\rangle } \right)
\end{equation}
and a matrix with next nonzero components
\[{\rm Z}_{14}  = \rho _0  ,\, {\rm
Z}_{26}  = {\rm Z}_{38}  =  - 2{\rm Z}_{45}  =  - 2{\rm Z}_{47}  =
2{\rm Z}_{49} = - 1/4\pi \rho _0  ,\]
\begin{equation}\label{12}  {\rm
Z}_{41}  = v_s^2 /\rho _0  ,\, {\rm Z}_{54}  ={\rm Z}_{74} = -
2{\rm Z}_{62}  = - 2{\rm Z}_{83}  = 2 \left\langle {E_0^2 }
\right\rangle /3.
\end{equation}
 In a plane  one-dimensional wave dependence
of vector of the state on a coordinate and time looks as
\cite{[2]} (p.49-55)
\begin{equation}\label{13} \Psi _\alpha = A_\alpha  \exp
\left( {ikz - i\omega t} \right).
\end{equation}
 Substitution of (\ref{13}) in
(\ref{10}) gives
\begin{equation}\label{14} {\rm Z}_{\alpha \beta }
A_\beta   = VA_\alpha,
\end{equation}
 where $V = \omega /k$
  is phase velocity of wave, $A_\alpha  $
  is a right eigenvector of matrix  ${\rm Z}$. Solving equation (\ref{14}) in standard way we find the
eigenvalues $V$
 of matrix ${\rm Z}$

\[ V = \left\{ {0,0,0, - \sqrt {\left\langle {E_0^2 } \right\rangle
 /12\pi \rho _0 } ,\sqrt {\left\langle {E_0^2 }
\right\rangle  /12\pi \rho _0 } , - \sqrt {\left\langle {E_0^2 }
\right\rangle  /12\pi \rho _0 } ,} \right.\]

\begin{equation} \left. \sqrt {\left\langle {E_0^2 } \right\rangle
/12\pi \rho _0 } , - \sqrt {v_s^2  + \left\langle {E_0^2 }
\right\rangle  /6\pi \rho _0 } ,\sqrt {v_s^2  + \left\langle
{E_0^2 } \right\rangle /6\pi \rho _0 }\right\} ,\label{15}
\end{equation}
and also eigenvectors. Because an eigenvector is determine within
a scalar multiplier, each of the found waves is characterized one
independent parameter as that we will choose amplitude of the
second central moment of the field. Non-spreading perturbations of
one of diagonal elements of the second central moment of the field
tensor and density correspond values   $V = 0$:
\[A_1  = \left\{
{1/8\pi v_s^2 ,0,0,0,0,0,0,0,1} \right\}, \,A_2  = \left\{ { -
1/8\pi v_s^2 ,0,0,0,0,0,1,0,0} \right\} ,\]
\begin{equation}\label{16} A_3  = \left\{ { - 1/8\pi v_s^2
,0,0,0,1,0,0,0,0} \right\}.
\end{equation}
 Fourth and fifth eigenvalues
correspond the mode of transversal velocity oscillations along an
axis  $x$
 and component  $\left\langle {E_x E_z } \right\rangle $:
 \[A_4  = \left\{ {0,\sqrt {3/4\pi \left\langle {E_0^2
} \right\rangle \rho _0 } ,0,0,0,1,0,0,0} \right\},\]
\begin{equation}\label{17} A_5  = \left\{ {0, - \sqrt {3/4\pi
 \left\langle {E_0^2 } \right\rangle \rho _0 }
,0,0,0,1,0,0,0} \right\}.
\end{equation}
 Sixth and seventh eigenvalues
correspond the mode of transversal velocity oscillations along an
axis $y$
  and component  $\left\langle {E_y E_z } \right\rangle $:
  \[A_6  = \left\{ {0,0,\sqrt {3/4\pi  \left\langle
{E_0^2 } \right\rangle \rho _0 } ,0,0,0,0,1,0} \right\},\]
\begin{equation}\label{18}
   A_7 =
\left\{ {0,0, - \sqrt {3/4\pi \left\langle {E_0^2 } \right\rangle
\rho _0 } ,0,0,0,0,1,0} \right\} .
\end{equation}
 Last two eigenvalues correspond the mode of longitudinal
velocity oscillations along an axis $z$, mass density and diagonal
components $\left\langle {E_x E_x } \right\rangle $
  and  $\left\langle {E_y E_y } \right\rangle $:
  \[A_8  = \left\{ {3\rho _0 /2 \left\langle {E_0^2 }
\right\rangle ,0,0, - 3\sqrt {v_s^2  + \left\langle {E_0^2 }
\right\rangle /6\pi \rho _0 } /2 \left\langle {E_0^2 }
\right\rangle ,1,0,1,0,0} \right\} ,\]
\begin{equation}\label{19}
A_9 = \left\{ {3\rho _0 /2 \left\langle {E_0^2 } \right\rangle
,0,0,3\sqrt {v_s^2  + \left\langle {E_0^2 } \right\rangle
 /6\pi \rho _0 } /2 \left\langle {E_0^2 }
\right\rangle ,1,0,1,0,0} \right\} .
\end{equation}
As it is obvious from (\ref{15})-(\ref{19}), there are two
transversal modes of oscillations with velocity

\begin{equation}\label{20}
v_t^{}  = \sqrt {\left\langle {E_0^2 } \right\rangle /12\pi \rho_0
}
\end{equation}
and one longitudinal with velocity  $ v_l^{}  = \sqrt {v_s^2  +
2v_t^2 } $. In this sense a situation is completely analogical to
the sound modes in an izotropic solid (see \cite{[3]} and
\cite{[6]} p. 124-128). If $ \left\langle {E_0^2 } \right\rangle
= 0 $, than we have ordinary sound in a liquid:  $ v_l^{}  =
v_s^{} $
 and  $
v_t^{}  = 0 $.
\section{Velocity of transversal sound in alkaline haloids}

We will apply an expounded  theory to the low-frequency long-wave
sound oscillations in an izotropic ionic crystal. We will study
alkaline haloids. As is generally known \cite{[7],[10]}, here
exactly electric interaction is decisive. Electrostatic potential
operating on a point ion is such (see \cite{[10]} (2.55)):
\begin{equation}\label{21} \varphi  =  - A {e}/{r},
\end{equation}
 here
$r$
 is distance between the nearest neighbours in a grate, $A$
 is the Madelung constant which is depending on crystal structure. We will take the structure of NaCl, for that  $A$
= 1.746 (see \cite{[10]} p. 65). Obviously, in a crystal there is
the microscopic electric field that is a random quantity with the
first moment of electric field strength $\vec E$
 is equal to zero. Exchange field is of the same order. We will suppose that  the second moment corresponds to (\ref{6b}).
 Characteristic strength of intracrystalline electric field proceed from (\ref{21}) is
 such $\left| {\vec E} \right| \sim \left| {Grad \varphi } \right| $,
 and its square can be estimated as

\begin{equation}\label{22}
\left\langle {E_0^2 } \right\rangle  = \left\langle {E_l E_l }
\right\rangle _0  \sim 4 \pi S {{e^2 }}/{{r^4 }} .
\end{equation}
A constant $S$ we can find from comparison with experimental data
once for NaCl structure.

 We use the tabular values of the bulk
modulus B (\cite{[7]} table 3.5) for the calculation of
transversal velocity of sound. It easily to obtain for the Young's
modulus Å expression through the bulk modulus B and  Poisson's
coefficient $ \mu $
 (see \cite{[6]} (5,9)) $
E = 3(1 - 2\mu )B $, then in mechanical approach (see \cite{[6]}
(22.4)) transversal velocity of sound is such

\begin{equation}\label{23}
v_t^{tab}  = \sqrt {\frac{{3(1 - 2\mu )B}}{{2\rho (1 + \mu )}}} .
\end{equation}
 On the other hand, according to the
formula (\ref{20}), it is possible to estimate the same velocity,
using electrohydrodynamic approach and estimation (\ref{22}). The
density of crystals is taken from \cite{[8]}, low-frequency
relative permittivity is taken from \cite{[7]} (table 5.1). The
values of Poisson's coefficient are taken from \cite{[9]}. Then
$S\approx 1.226$. Values of relative permittivity, marked with the
apostrophe ', are taken from \cite{[10]} (table 2.2). $r$
 is taken from \cite{[7]} (table 3.5). All data are at normal conditions.

\begin{table}[a]
\noindent\caption{Velocity of transversal sound in some alkaline
haloids}\vskip3mm\tabcolsep4.2pt

\noindent{\footnotesize
\begin{tabular}{c c c  c c c c}
 \hline%
 \multicolumn{1}{c}{\rule{0pt}{9pt}crystal}%
 & \multicolumn{1}{|c}{ $ B \,10^{11} dyn / sm^{
 3} $} & \multicolumn{1}{|c}{ $\mu $} & \multicolumn{1}{|c}{ $
\rho _0 \,g / sm^{  3} $} & \multicolumn{1}{|c}{ $ r\,\mathop {\rm
A}\limits^ \circ $}
 & \multicolumn{1}{|c}{ $
v_t^{tab} \,10^5 \,sm/s $}
 & \multicolumn{1}{|c}{ $
v_t \,10^5 \,sm/s
$}\\%
\hline%

\rule{0pt}{9pt}

LiF&6.71&0.214&2.64&2.014&4.24&4.66
\\
LiCl&2.98&0.245&2.07&2.570&2.97&3.23
\\
LiBr&2.38&0.256&3.46&2.751&2.00&2.18
\\
NaF &4.65  &  0.234&   2.56&   2.317   &3.43&    3.57
\\NaCl& 2.40& 0.243&   2.17&    2.820 &  2.62  &  2.62 \\NaBr
& 1.99& 0.270& 3.21&  2.989   &1.84 &   1.92 \\KF&  3.05 & 0.274&
2.48&  2.674& 2.56 &2.72 \\KCl& 1.74&    0.259&   1.98& 3.147
&2.25& 2.20
\\KBr& 1.48 &   0.283   &2.75&     3.298  & 1.65    &1.70 \\KI&
1.17& 0.265 &3.13 &  3.533&   1.44&    1.39 \\RbF& 2.62& 0.276&
2.88&  2.815& 2.19   & 2.28 \\RbCl    &1.56   & 0.268 &2.76 &
3.291& 1.76& 1.71 \\RbBr  &  1.30   & 0.267   &2.78 & 3.445 &1.61&
1.55 \\RbI& 1.06& 0.309   &3.56&    3.671 &1.14& 1.21

\\\hline

\end{tabular}
}\label{tabl}
\end{table}

As we see in table \ref{tabl}, values of transversal velocity that
it is necessary to equate at a velocity of sound, calculated on
the obtained formula (\ref{20}), even at so rough estimation of
equilibrium correlation value of the electrostatic field according
to a formula (\ref{22}), surprisingly well coincide with
calculated, proceed from these mechanical data of crystals, on a
formula (\ref{23}).


\section{Conclusions}
\begin{itemize}
\item  Thus, the evolution of dielectric with a constant relative permittivity is studied in approximation of ideal
hydrodynamics. Electric field as a random quantity is
characterized by a nonzero equilibrium value of the second central
moment of electric field strength. The linear system of equations
is obtained for the mass density, velocity and tensor of the
second central moment of electric field strength.

\item  Adiabatic modes in this system are studied. Found two transversal with consilient phase velocities and
longitudinal sound modes.
Converting in zero the equilibrium value of the second central
moment of electric field strength transversal modes disappear, and
a longitudinal mode passes to the ordinary sound in a liquid.

\item The estimation of characteristic equilibrium value of the second central moment of electric field strength
for an ionic crystal is produced, that showed good accordance of the values calculated proceed from expounded an
electrohydrodynamic theory of transversal sound velocity to the values that is given by an ordinary mechanical theory.

\end{itemize}



\begin{thebibliography}{10}
\bibitem{[1]}{\sc A.I.~Zhakin}. \newblock{Electrohydrodynamics.} {\it Phys. Usp. }, vol.~55 (2012),
pp.~465–-488.
\bibitem{[3]}{\sc A.A.~Stupka}. \newblock{Hydrodynamic modes of conducting liquid in accidental
magnetic field.} {\it  Magnetohydrodynamics}, vol.~2 (2010),
pp.~137--141.
\bibitem{[4]} {\sc A.A.~Stupka}. Sound modes in isotropic magnetic
hydrodynamics. {\it The Journal of Kharkiv National University.
Physical series: Nuclei, Particles, Fields}, vol.~859 (2009),
no.~2(42). pp.~56--58. (in Russ.).
\bibitem{[5]} {\sc Elektrodinamika plazmy}.  Edited by A.I.~Akhiezer (Nauka, Moscow, 1974)
\newblock {(in Russ)}.

\bibitem{[11]}
{\sc L.D.~Landau, E.M.~Lifshitz and L.P.~Pitaevskii}.  {\it
Electrodynamics of Continuous Media.} (Vol. 8 (1rst ed.),
Butterworth-–Heinemann, 1984).
\bibitem{[2]} {\sc R.V.~Polovin and V.P. Demutskiy}.
{\it  Osnovy magnitnoy gidrodinamiki} (in Russ.) (Energoatomizdat,
Moscow,  1987).



\bibitem{[6]}{\sc L.D.~Landau and E.M.~Lifshits}.  {\it  Theory of Elasticity.}
 (Vol. 7 (3rd ed.). Butterworth–-Heinemann,
1986).
\bibitem{[7]} {\sc  Charles Kittel}. {\it  Introduction to Solid State Physics}. (John Wiley
and Sons LTD,  2004).
\bibitem{[10]}{\sc J.A.~Reissland}. {\it The physics of
phonons} (John Wiley and Sons LTD, London - New York - Sydney -
Toronto, 1973).
\bibitem{[8]} {\sc B.A.~Rabinovich and Z.Ya.~Havin}  {\it  Kratkiy himicheskiy spravochnik}
 (Himiya, Leningrad, 1978) (in Russ.).
\bibitem{[9]}{\sc V.N.~Belomestnykh and E.P.~Tesleva}.
\newblock{ Interrelation
between anharmonicity and lateral strain in quasi-isotropic
polycrystalline solids.} {\it Technical Physics}, vol.~49 (2004),
no.~8,  pp.~1098--1100.







\end{thebibliography}


\end{document}